\documentclass[sigconf]{acmart}
\usepackage{siunitx}

\renewcommand\footnotetextcopyrightpermission[1]{}

\AtBeginDocument{%
  \providecommand\BibTeX{{%
    \normalfont B\kern-0.5em{\scshape i\kern-0.25em b}\kern-0.8em\TeX}}}

\setcopyright{none}
\copyrightyear{2021}
\acmYear{2021}
\acmDOI{}

\acmConference{}
\acmBooktitle{}
\acmPrice{}
\acmISBN{}

\begin{document}

\title{Deep Reinforcement Learning based Group Recommender System}


\author{Zefang Liu, Shuran Wen, Yinzhu Quan}
\authornote{All authors contributed equally to this research.}
\affiliation{%
  \institution{Georgia Institute of Technology}
  \city{Atlanta}
  \state{Georgia}
  \country{USA}
}
\email{{liuzefang,swen43,yquan9}@gatech.edu}


\begin{abstract}


Group recommender systems are widely used in current web applications. In this paper, we propose a novel group recommender system based on the deep reinforcement learning. We introduce the MovieLens data at first and generate one random group dataset, MovieLens-Rand, from it. This randomly generated dataset is described and analyzed. We also present experimental settings and two state-of-art baselines, AGREE and GroupIM. The framework of our novel model, the Deep Reinforcement learning based Group Recommender system (DRGR), is proposed. Actor-critic networks are implemented with the deep deterministic policy gradient algorithm. The DRGR model is applied on the MovieLens-Rand dataset with two baselines. Compared with baselines, we conclude that DRGR performs better than GroupIM due to long interaction histories but worse than AGREE because of the self-attention mechanism. We express advantages and shortcomings of DRGR and also give future improvement directions at the end.

\end{abstract}





\maketitle

\pagestyle{plain}
\section{Introduction}

Nowadays, Internet is essential in our daily life. People use their electronic products such as smart phones, laptops at all time. The rapid growth of online platforms and applications has provided a great many choices and services. During the pandemic, people could stay at home and get entertainment from the Internet. Under these circumstances, recommender system plays an important role to recommend stuff like movies, news, songs, and games according to users preferences in order to save people's searching time. In many scenarios, recommendations are needed for a group of users. As a result, our aim is to implement a novel group recommender system to give the most suitable recommendations for groups.

With a lot of fantastic deep learning recommendation models, such as Deep Crossing \cite{ying2016deep} and Wide \& Deep \cite{cheng2016wide} for individuals, MoSAN \cite{tran2019interact} and GroupIM \cite{sankar2020groupim} for groups, recommender system enters the era of deep learning in an all-round way. Compared to traditional machine learning models, deep learning models are more flexible and expressive. However, existing group recommendation models meet three challenges as shown below and can not address them in most efficient ways. Firstly, it is difficult to deal with the dynamic changes in recommendations. It is mainly because user preferences update very quickly nowadays, and users' interests can change as time goes by. Secondly, existing group recommender systems can only optimize current (short-term) rewards rather than future (long-term) ones, which are more important to user experiences and company profits. Lastly, current group recommendation models always recommend items based on one same group profiles and ignore relationship among diverse items, which might make users a little bit bored. Therefore, it is better to add some randomness for diversity. 

In order to overcome these limitations of current group recommender systems, we propose one novel Deep Reinforcement learning based Group Recommender system (DRGR)\footnote{GitHub: \url{https://github.com/zefang-liu/group-recommender}}. We model the group recommendation task as one Markov Decision Process (MDP), where state and action spaces are defined accordingly. One environment simulator based on the matrix factorization is built to simulate this MDP. The agent of the DRGR is constructed with actor-critic networks and is optimized with the Deep Deterministic Policy Gradient (DDPG) algorithm. This novel model is implemented on one randomly generated group dataset MovieLens-Rand from the original MovieLens dataset with two state-of-art baselines, AGREE and GroupIM. The DRGR over-performances one baseline GroupIM over the evaluation metrics of recall and Normalized Discounted Cumulative Gain (NDCG).

Our model DRGR is meaningful to those people who want to get recommendations for their groups, such as entertainments with families and travels with friends. The DRGR could help groups select their favorite items according to their common preferences and save their valuable time in the most effective way.

\section{Literature Survey}

Deep reinforcement learning (DRL) has been introduced into recommender systems for individuals \cite{zhang2019deep,zheng2018drn,zhao2019deep,zhao2018deep,zhao2018recommendations,liu2019deep,chen2018stabilizing,choi2018reinforcement,munemasa2018deep,wang2014exploration}. Zhao et al. \cite{zhao2019deep} proposed a list-wise recommendation framework based on deep reinforcement learning, LIRD, for a real-world e-commerce environment. The recommendation session is viewed as a Markov Decision Process (MDP), an online user-agent interacting environment simulator is designed, and an Advantage Actor-Critic (A2C) model is trained with Deep Deterministic Policy Gradient (DDPG) algorithm. LIRD outperforms Deep Neural Network (DNN) and Recurrent Neural Network (RNN) in Mean Average Precision (MAP) and Normalized Discounted Cumulative Gain (NDCG), also Deep Q-Network (DQN) in speed. This model will be generalized to group recommendation in our work. For shortcomings, the cosines similarity is used in this work to get the reward for an unknown state-action pair in the memory, which can be replaced by nonlinear functions such as neural networks. 

Tran et al. \cite{tran2019interact} proposed a Medley of Sub-Attention Networks (MoSAN) for the ad-hoc group recommendation. This model can dynamically learn user impact weights for groups and user-user interactions in one group. Besides, it can provide the relative importance of users in a group for explainable recommendations. MoSAN overperforms state-of-the-art baselines on four real-world datasets with a considerable margin. The attention network structure used in this model from the representation learning can be used for group embeddings in our work. However, the limitation of MoSAN is that temporal orders in the group preferences are ignored, and hence MoSAN overlooks the evolution of group interests with time.

Sankar et al. \cite{sankar2020groupim} proposed a recommender architecture-agnostic framework, named Group Information Maximization (GroupIM), for item recommendations to ephemeral groups with limited historical activities together. Data-driven regularization strategies including contrastive representative learning and group-adaptive preference prioritization are proposed to exploit the preference covariance and contextual relevance. GroupIM overperformances state-of-art group recommenders on several real-world datasets by Recall@K and NDCG@K. This model provides a high-level framework for group recommendation and will be used as a baseline for our work. Also, the idea of data-driven user preference aggregations can be used in our project for user state aggregations. Limitations of this work include parameter tuning in the overall strength of preference regularization and missing of side information or contextual factors. Besides, the reinforcement learning can be applied to adapt to the temporal dynamics of the group recommender.

\section{Dataset Description and Analysis}

In this section, we first describe the data preparation process, then do the raw data statistics and data analysis.

\subsection{Data Preparation}

We are going to use the MovieLens \cite{harper2015movielens} dataset, which describes 5-star rating and free-text tagging activity from MovieLens, a movie recommendation service. The dataset can be download from \href
{https://grouplens.org/datasets/movielens/}
{MovieLens website}\footnote{\url{https://grouplens.org/datasets/movielens/}}.
For data preprocessing, we firstly randomly generate groups with 2-5 users. Then, for each group, if every member gives 4-5 stars to one movie, we assume that this movie is adopted by this group with rating 1. If all members give ratings to one movie, but not all in 4-5 stars, we consider the group gives rating 0 to this movie. For other cases, the group movie ratings are missed. Finally, to ensure each group has enough interactions with items, we require each group has at least 20 ratings. Also, for each rating, 100 rating-missed items are randomly sampled.

\subsection{Raw Data Statistics}

\begin{figure}[!h]
    \centering
    \includegraphics[width=\linewidth]{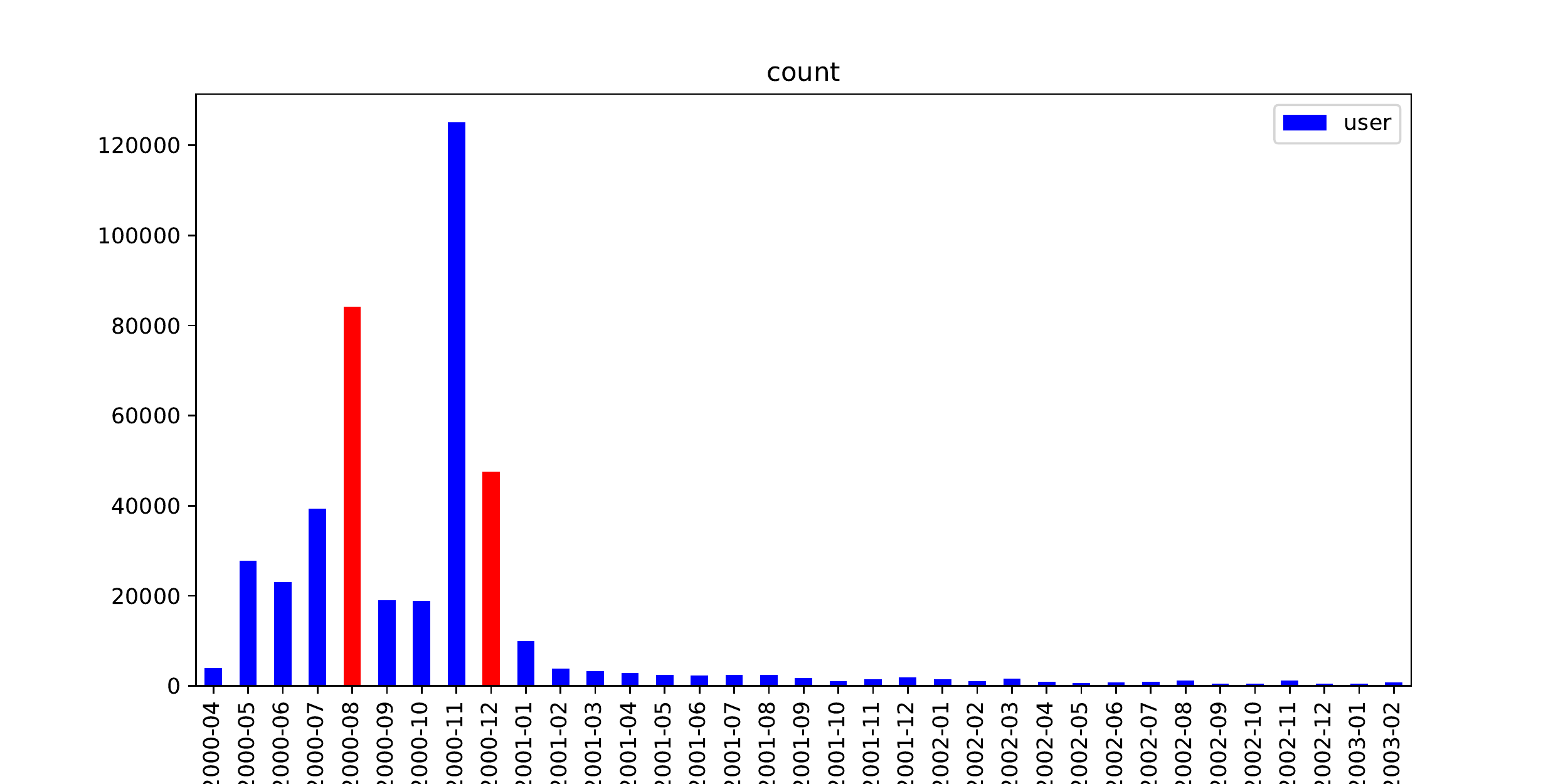}
    \caption{Number of ratings for each month counts, where rad bars are data splitting.}
    \label{fig:date}
\end{figure}

After applying the data pre-processing strategy to the original dataset, we obtain the new \texttt{MovieLens-Rand} dataset. It consists of generally three types of data: group member, group rating, and user rating. Group member is made up of groupId and userIds. Group rating is made up of groupId, movieId, rating, and timestamp. User rating is made up of userId, movieId, rating, and timestamp. The raw data statistics of the new dataset are shown in the Table \ref{tab:data}, where the number of users, items, and groups are listed with some statistics for rating numbers. The average number of ratings per group is about 53, which is large enough to train baselines (AGREE and GroupIM) and our DRGR model. Also, the Figure \ref{fig:date} shows the count of ratings per month. This give us information about how the data is split, since our reinforcement method needs to take timestamp into account. The two red bars in the Figure \ref{fig:date} split the date ranges into the training, validation, and testing parts. 

\begin{table}[!h]
\centering
\caption{Summary statistics of the dataset, where U-I and G-I stand for user-item and group-item respectively.}
\label{tab:data}
\begin{tabular}{c c} 
    \toprule
    Dataset & MovieLens-Rand \\
    \midrule
    \# Users & 1626 \\
    \# Items & 1998 \\
    \# Groups & 1000 \\
    \# U-I ratings & \SI{438129}{} \\
    \# G-I ratings & \SI{53248}{} \\
    Avg. \# ratings / user & 269.45 \\
    Avg. \# ratings / group & 53.25 \\
    Avg. group size & 2.19 \\
    \bottomrule
\end{tabular}
\end{table}

\subsection{Data Analysis}

\begin{figure}[!h]
    \centering
    \includegraphics[width=\linewidth]{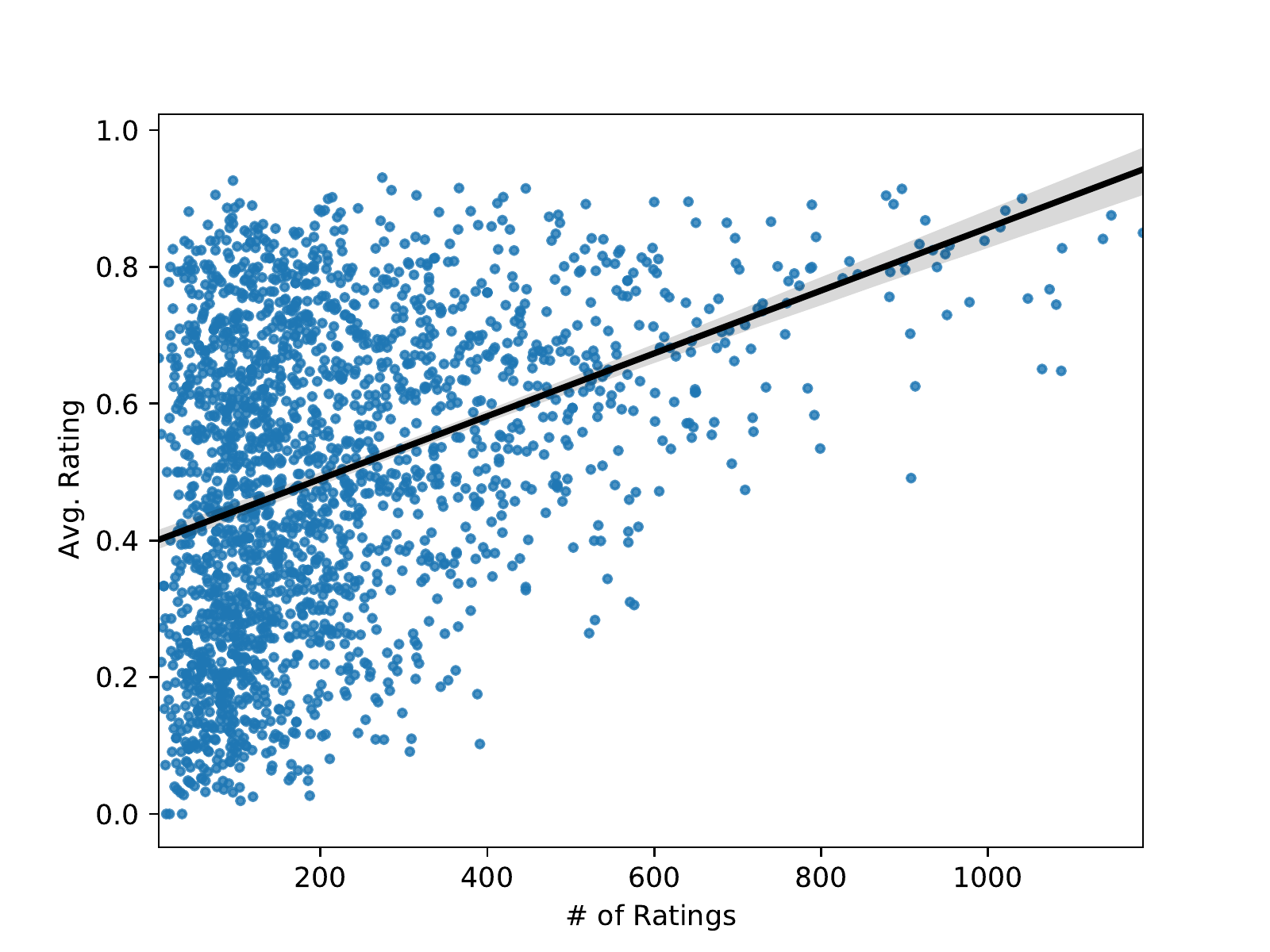}
    \caption{The average rating versus the number of ratings for movies, where the black line shows the liner regression.}
    \label{fig:num_ratings}
\end{figure}

\begin{figure}[!h]
    \centering
    \includegraphics[width=\linewidth]{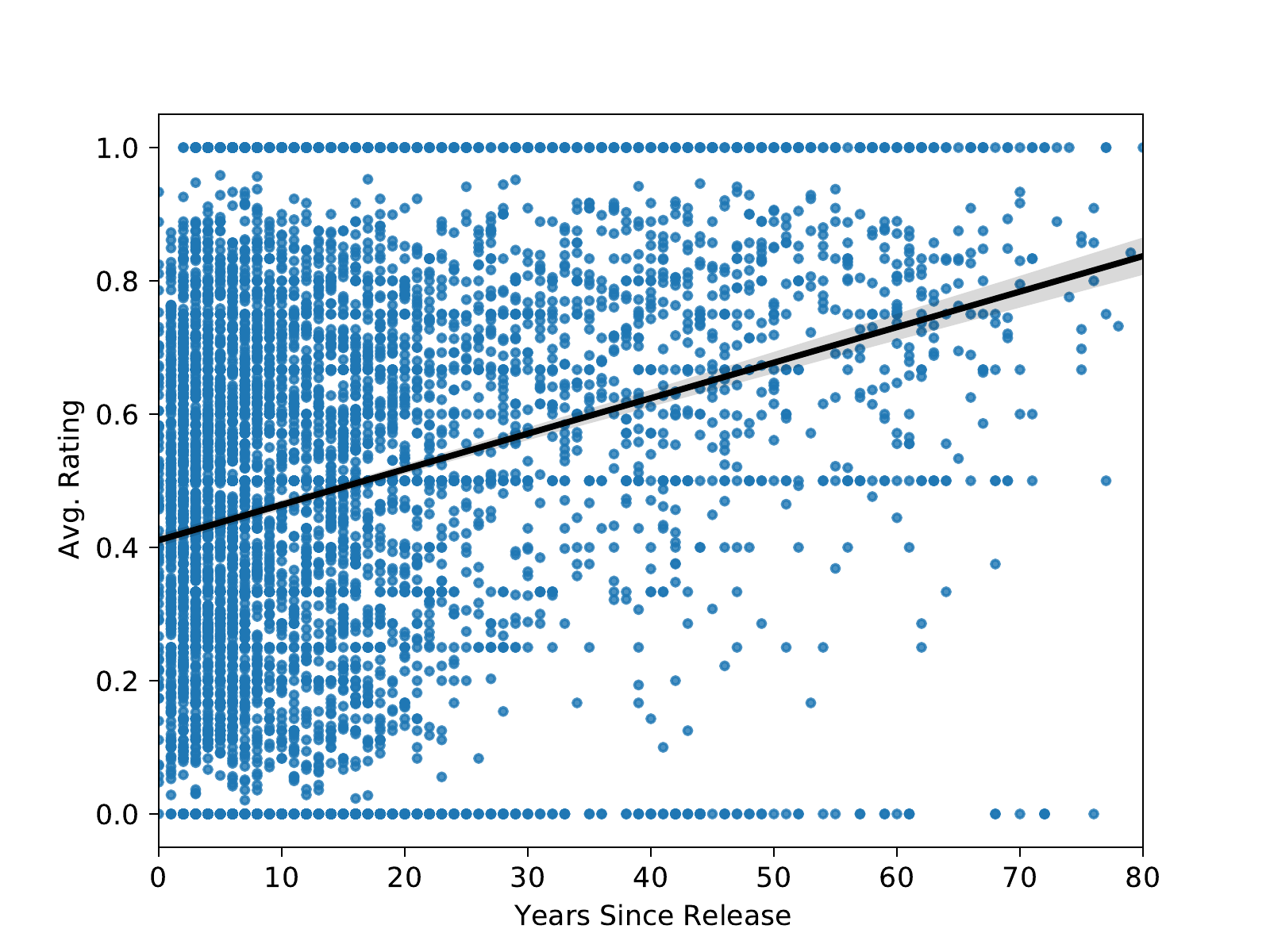}
    \caption{The average rating versus the years since movie release, where the black line shows the liner regression.}
    \label{fig:years_since_release}
\end{figure}

\begin{figure}[!h]
    \centering
    \includegraphics[width=\linewidth]{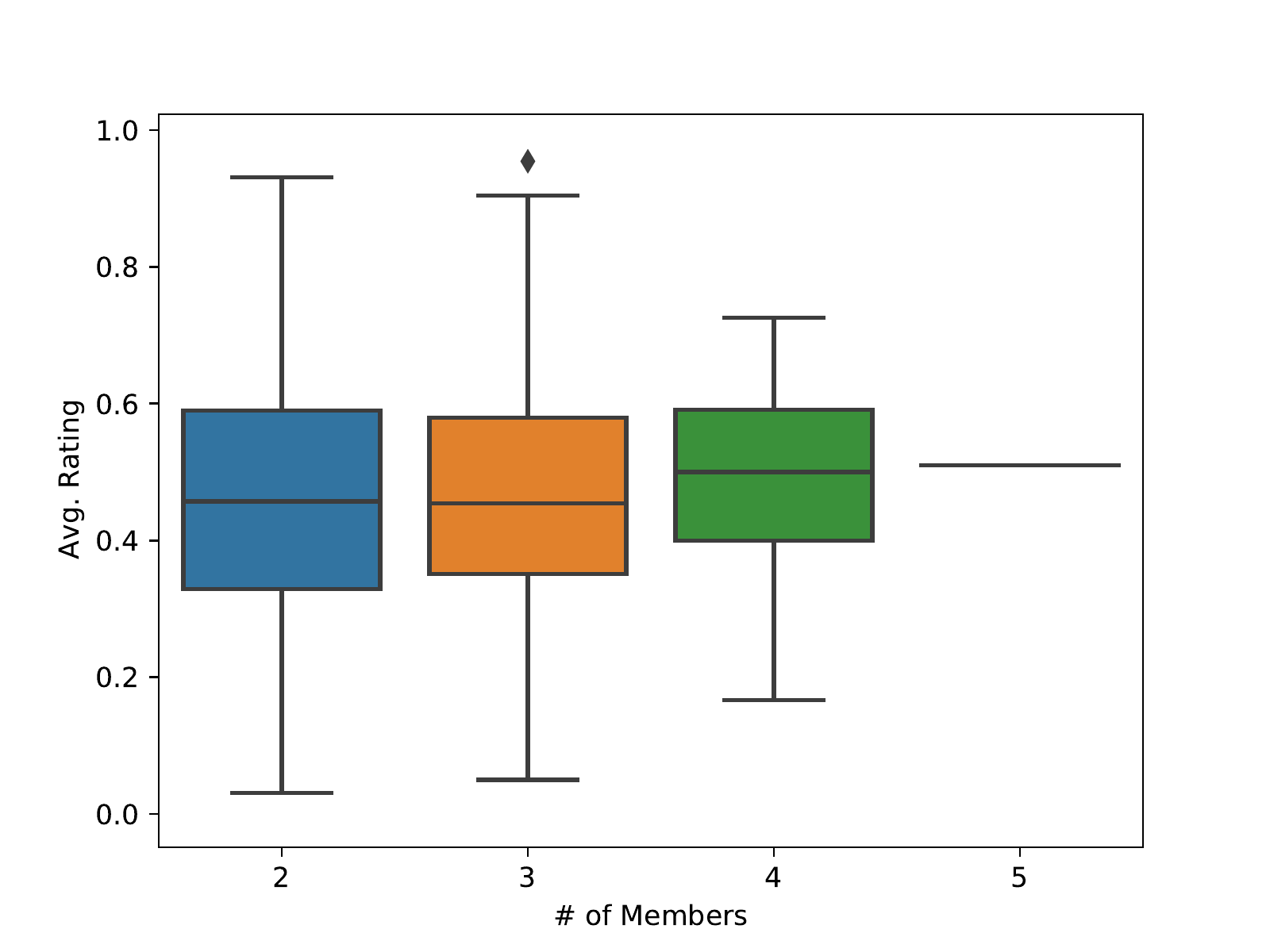}
    \caption{Average ratings for different group sizes.}
    \label{fig:num_members}
\end{figure}

We describe the data analysis for the \texttt{MovieLens-Rand} dataset in this part. The Figure \ref{fig:num_ratings} shows the distribution of the average user ratings for all the movies. The average ratings is found to increase with the number of ratings. The Figure \ref{fig:years_since_release} shows how average ratings change with the years since movie release. We find the average rating of one movie increases with the year of release, which represents the dynamics of ratings. The Figure \ref{fig:num_members} shows the distribution of average ratings for different group sizes. Although the average ratings does not change with the group sizes, the interquartile range (IQR) becomes smaller for the larger group, which shows the larger group can balance member interests. 

\section{Experimental Settings and Baselines}

In this section, we present experimental settings and briefly introduce two baselines we use.

\subsection{Experimental Settings}

The group recommendation is viewed as a classification task. When one item is recommended to a group, if the group chooses the item, this case is marked as a positive sample. Otherwise, it will be a negative sample. 

The evaluation metrics used in this paper are recall (RECALL@K) and normalized discounted cumulative gain (NDCG@K), where $ K $ is the number of recommendations and $ K = \{ 5, 10, 20 \} $. The RECALL@K is the fraction of relevant items that have been retrieved in the top K recommended items, and NDCG@K takes account the item rankings in the recommendation list. All metrics with higher values indicate a better model. These metrics are widely used in evaluating recommender systems \cite{cao2018attentive,tran2019interact,sankar2020groupim}.

Both user and group rating data are split into training, validation, and testing datasets with the ratio of 70\%, 10\%, and 20\%  respectively by the temporal order. All models are trained on the training set, the hyper-parameters are tuned on the validation set, and the evaluation of models is done on the testing set. The codes are run on the Google Colab \footnote{\url{https://colab.research.google.com/}}, where one Tesla P100-PCIE-16GB GPU is used. The Python version is 3.7.10, and the PyTorch \cite{adam2019pytorch} version is 1.8.1.

\subsection{Baseline Description}

With the purpose of evaluations for our model performances on the \texttt{MovieLens-Rand} dataset, we use two state-of-art baselines.

The first baseline is the Attentive Group Recommendation (AGREE) \cite{cao2018attentive} model, and its GitHub repository is Attentive-Group-Recommendation\footnote{\url{https://github.com/LianHaiMiao/Attentive-Group-Recommendation}}. In the representation layer of this model, an attention mechanism is adopted to represent groups, where group members and items are embedded first, then sent to a neural attention network to get the group embeddings. Later, these group embeddings are stacked with item embeddings in the pooling layer, then sent to hidden layers and the following prediction layer. The neural collaborative filtering (NCF) is used here to learn the group/user-item interactions. Hence, this model can recommend items for both groups and users. The hyper-parameters are shown in the Table \ref{tab:hyper-parameters}, which is tuned on the validation dataset. The AGREE model is one of the recent group recommendation models with good performances on the real-world datasets, which is suitable to be compared with our group recommendation model.

The second baseline is the Group Information Maximization (GroupIM) \cite{sankar2020groupim} model, and its Github repository is GroupIM\footnote{\url{https://github.com/CrowdDynamicsLab/GroupIM}}. In this model, data-driven
regularization strategies are proposed to exploit both the preference covariance amongst users who are in the same group, as well as the contextual relevance of users’ individual preferences to each group. GroupIM model made two contributions. First, the recommender
architecture-agnostic framework GroupIM can integrate arbitrary neural preference encoders and aggregators for ephemeral group recommendation. Second, it regularizes the user-group latent space to overcome group interaction sparsity by: maximizing mutual information between representations of groups and group members; and dynamically prioritizing the preferences of highly informative members through contextual preference weighting. The hyper-parameters are shown in the Table \ref{tab:hyper-parameters}, which is tuned on the validation dataset. The GroupIM experimental results on several real-world datasets indicate significant performance improvements over state-of-the-art group recommendation techniques.

\begin{table}[h!]
\centering
\caption{Hyper-parameters for AGREE, GroupIM, and DRGR models, where LR stands for the learning rate, the number of LR decay steps means when the LR multiplies the LR decay ratio per LR decay steps, and MF is for matrix factorization.}
\label{tab:hyper-parameters}
\begin{tabular}{c c c c} 
    \toprule
    Model & AGREE & GroupIM & DRGR\\
    \midrule
    Embedding size & 32 & 32 & 32 \\ 
    Epoch & 15 & 20 & 1000 \\
    Batch size & 256 & 256 & 64\\
    Negative samples & 100 & - & 100 \\
    Initial LR & \SI{0.00001}{} & 0.005 & \SI{0.0001}{}\\ 
    LR decay ratio & 0.5 & 0.4 & -\\ 
    LR decay steps & 10 & - & -\\
    Weight decay & - & - & \SI{E-6}{}\\
    \midrule
    History length & - & - & 5 \\
    MF component num. & - & - & 32 \\
    Actor hidden sizes & - & - & (128, 64)\\
    Critic hidden sizes & - & - & (32, 16)\\
    $ \gamma $ & - & - & 0.9\\
    \bottomrule
\end{tabular}
\end{table}

\section{Proposed Methods}

In this section, we present the problem statement and the algorithm framework for our Deep Reinforcement Group Recommendation (DRGR) model. The Markov Decision Process is introduced for this group recommendation task, and the task is resolved by the deep reinforcement learning. We will also discuss the novelties of our method.

\subsection{Problem Statement}

In this work, the recommender system is viewed as an agent, which interacts with the environment $ \mathcal{E} $ by recommending items to maximize its reward. The environment is a set of multiple groups, where one single user is assumed to be one member group here. This task is modeled as a Markov Decision Process (MDP). Formally, this MDP contains five elements $ (\mathcal{S}, \mathcal{A}, \mathcal{P}, \mathcal{R}, \gamma) $ as follows. Let $ \mathcal{U} $, $ \mathcal{G} $, and $ \mathcal{I} $ be sets of user, group, and item ids respectively.

\begin{itemize}
    \item State space $ \mathcal{S} $: A state $ s_t \in \mathcal{S} $ represents the state of a group at time $ t $. A state $ s_t = [g, h_t] $ consists two parts, one for the group id $ g \in \mathcal{G} $ and another for the group browsing history $ h_t $. The group id $ g $ can be mapped to its members, i.e. $ g = \{ u_1, u_2, \dots \} $, where $ u_1, u_2, \dots \in \mathcal{U} $. The browsing history of a group is $ h_t = [i_1, i_2, \dots, i_N] $, where $ i_1, i_2, \dots, i_N \in \mathcal{I} $.
    
    \item Action space $ \mathcal{A} $: An action $ a_t = [i_{t,1} , i_{t,2} , \dots, i_{t,K}] \in \mathcal{A} $ is a list of items recommended to a group from the recommender system, where $ K $ is the number of items. For learning efficiency, we will use $ K = 1 $ when the agent interacts with the environment, i.e. $ a_t \in \mathcal{I} $.
    
    \item Reward $ \mathcal{R} $: After an action $ a_t $ is taken by the recommender to a group at one state $ s_t $, the recommender will receive a reward $ r_t \in \{0, 1 \} $ based on the group response. The reward will be $ r_t = 1 $, if the recommended item is picked by this group. Otherwise, $ r_t = 0 $.
    
    \item Transition probability $ \mathcal{P} $: Transition probability is defined as $ p(s_{t+1}| s_{t}, a_{t}) $, which satisfies the Markov property. This probability measures how the environment evolve with the time $ t $.
    
    \item Discount factor $ \gamma $: The $ \gamma \in [0,1] $ measures how the future reward will be valuated today.
\end{itemize}
By given this MDP $ (\mathcal{S}, \mathcal{A}, \mathcal{P}, \mathcal{R}, \gamma) $, the agent will try to maximize its rewards by taking actions to interact with the environment. The solution will be one policy $ \pi : \mathcal{S} \to \mathcal{A} $. 

\subsection{Environment Simulator}

\begin{figure}[!h]
    \centering
    \includegraphics[width=.8\linewidth]{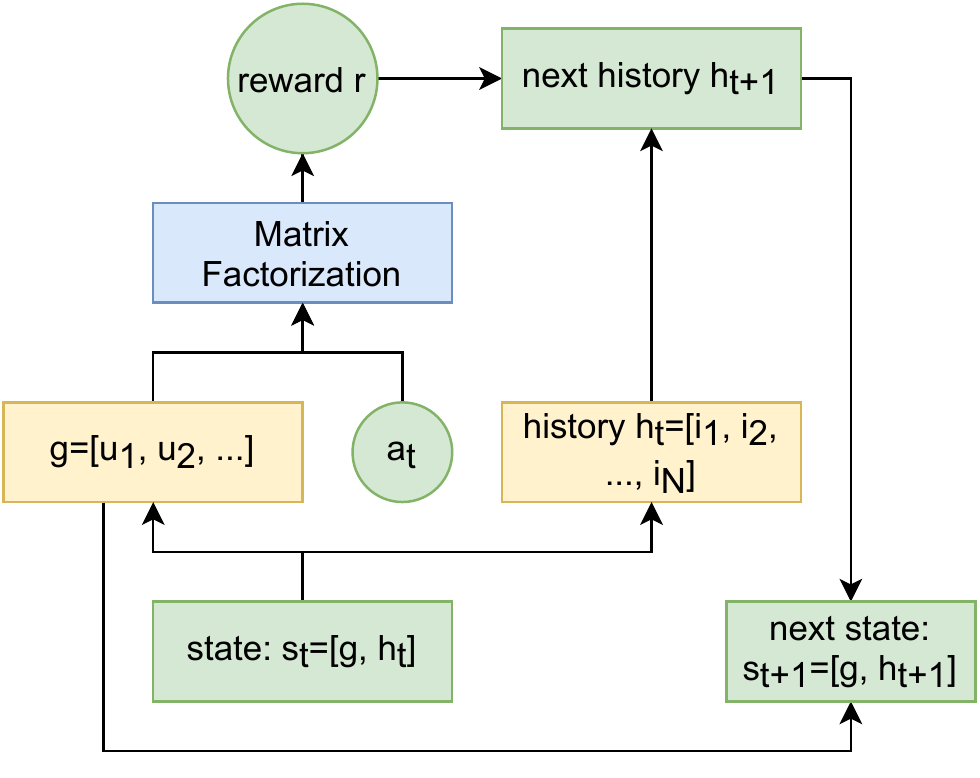}
    \caption{The framework of the environment, where the matrix factorization is used to simulate unknown rewards.}
    \label{fig:env}
\end{figure}

Since the rating matrix of group-item pairs is sparse, one environment simulator is needed to simulate this MDP, which is shown in the Figure \ref{fig:env}. The matrix factorization is used to predict the unknown ratings for the state $ s_t $ with the action $ a_t $, which will be used as the reward $ r_t $. If $ r_t > 0 $, $ h_{t+1} = [i_2, \dots, i_N, a_t] $. Otherwise, $ h_{t+1} = h_t $. The next state will be $ s_{t+1} = [g, h_{t+1}] $.

\subsection{Agent Framework}

\begin{figure}[!h]
    \centering
    \includegraphics[width=\linewidth]{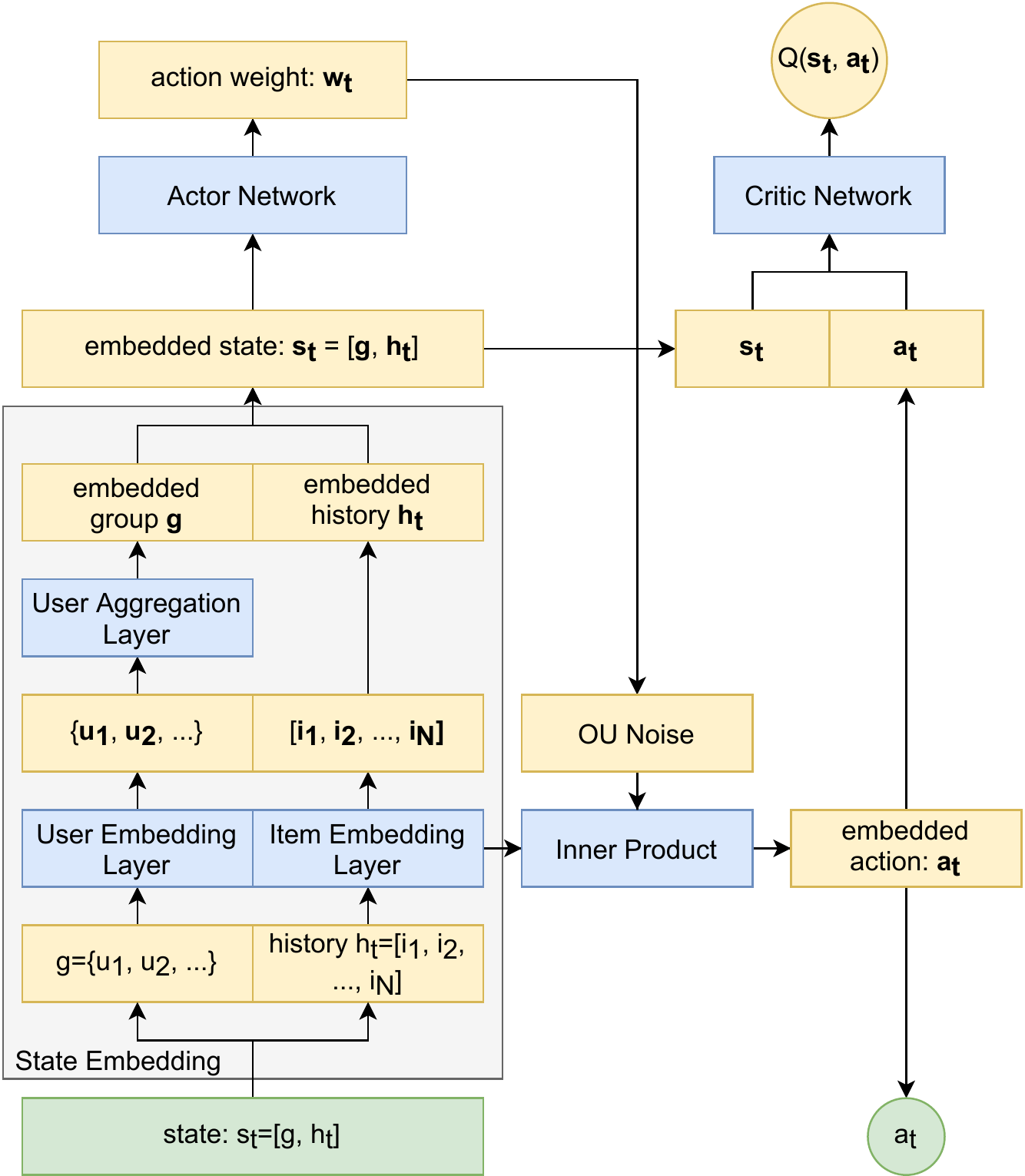}
    \caption{The framework of the agent, where state embedding layer, actor network, and critic network are included.}
    \label{fig:agent}
\end{figure}

In order to solve this MDP problem, one agent based one the actor-critic model is built, which is shown in the Figure \ref{fig:agent}. This agent model consists three networks, i.e. state embedding network, actor network, and critic network. 
\begin{itemize}
    \item \textbf{State embedding}: The state embedding network is to embed one state $ s_t = [g, h_t] $ to its embedding $ \mathbf{s}_t $. The group id $ g $ is mapped to its group members $ g = \{ u_1, u_2, \dots \} $ first, then those user ids are embedded through the user embedding layer to $ \{ \mathbf{u}_1, \mathbf{u}_2, \dots \} $. One self-attention user aggregation layer is used, i.e. 
    \[
    o_j = \mathbf{h}^{\text{T}} \text{ReLU} (\mathbf{P} \mathbf{u}_j + \mathbf{b}) , \quad \alpha_j = \text{softmax}_{u_j \in g} (o_j) ,
    \]
    \[
    \mathbf{g} = \sum_{u_j \in g} \alpha_j \mathbf{u}_j ,
    \]
    to get the embedded group $ \mathbf{g} $. Meanwhile, the history $ h_t =[i_1, i_2, \dots, i_N] $ is embedded through the item embedding layer to the embedded history $  \mathbf{h}_t = [\mathbf{i}_1, \mathbf{i}_2, \dots, \mathbf{i}_N] $. Combining these two parts, we get the embedded state $ \mathbf{s}_t = [\mathbf{g}, \mathbf{h}_t] $.
    
    \item \textbf{Actor}: The actor is a multi-layer neural network, whose input is the embedded state $ \mathbf{s}_t $ and output is the action weight $ \mathbf{w}_t $. To promote exploration, one Ornstein–Uhlenbeck process noise can be added here. This action weight will take inner products with item embeddings, i.e. $ s_j = \mathbf{w}_t^{\text{T}} \mathbf{i}_j $. The item with the highest $ s_j $ will be recommended to the group, and the corresponding item embedding will be sent to the critic next. The actor network can be optimized by the policy gradient.
    
    \item \textbf{Critic}: The critic is also a multi-layer neural network, which will assign the Q-value of one state-action pair, i.e. $ Q(\mathbf{s}_t, \mathbf{a}_t) $. The critic network can be optimized by the temporal difference method as the Deep Q Learning.
\end{itemize}
The deep deterministic policy gradient (DDPG) algorithm is used to train parameters in the agent, where one experience reply and target networks are used. The kernels and hyper-parameters of the model have been shown in the Table \ref{tab:hyper-parameters}.

\subsection{Novelties of the Algorithm}

Several shortcoming of the previous group recommender systems are overcame. Compared with LIRD \cite{zhao2019deep}, we generalize the DRL framework to the group recommendation task. Instead of viewing a group as one user, we consider the influences of each group member by one self-attention mechanism. Compared with AGREE \cite{cao2018attentive} and GroupIM \cite{sankar2020groupim}, by formulating the group recommendation as a MDP, the dynamics property of the task can be learned, where the temporal structure of the data can play an important role. This improved algorithm allows our DRGR to achieve comparable results over previous group recommender system baselines \cite{cao2018attentive,sankar2020groupim}.

\section{Experiments and Results}


We implement two baselines, AGREE and GroupIM, with our DRGR model on the \texttt{MovieLens-Rand} dataset. After trained on the training set and tuned on the validation set, the model results on the testing set are shown in the Table \ref{tab:baselines}.

\begin{table}[h!]
\centering
\caption{Group recommendation results of AGREE, GroupIM, and DRGR, on the \texttt{MovieLens-Rand} dataset, where R@K and N@K denote the Recall@K and NDCG@K metrics at K = 5, 10 and 20.}
\label{tab:baselines}
\begin{tabular}{c | c c c | c c c} 
    \toprule
    Dataset & \multicolumn{6}{c}{MovieLens-Rand} \\
    Metric & R@5 & R@10 & R@20 & N@5 & N@10 & N@20 \\
    \midrule
    AGREE & \textbf{0.4006} & \textbf{0.5591} & \textbf{0.7335} & \textbf{0.2737} & \textbf{0.3250} & \textbf{0.3691} \\
    GroupIM & 0.1576 & 0.1688 & 0.2497 & 0.1597 & 0.1602 & 0.1891 \\
    DRGR & 0.2885 & 0.4328 & 0.5572 & 0.1874 & 0.2336 & 0.2653 \\
    \bottomrule
\end{tabular}
\end{table}

By viewing results on the \texttt{MovieLens-Rand} dataset, we find that DRGR performs better than GroupIM but worse than AGREE on recall and NDCG metrics. From the GroupIM model, since GroupIM targets ephemeral groups lack of historical interactions, it contributes a lot in a situation with group interaction sparsity. However, the average number of group ratings is 53.25 in the \texttt{MovieLens-Rand} dataset, which is relatively large. The reason causes that DRGR overcomes GroupIM. For the AGREE model, AGREE has a neural attention mechanism, which can learn influences of group members and adapt their influences for different items. However, Our DRGR model uses one self-attention mechanism, which can only aggregate group member preferences but ignore their interactions with different items in the potential recommendation list. This makes our model inflexible when item kinds are diverse. This drawback leads that AGREE performs better then our DRGR model under this circumstance.

\section{Conclusion}

In summary, by viewing the group recommendation task as a Markov decision process, the Deep Reinforcement learning based Group Recommender system (DRGR) makes use of the group historical data and keeps updating the model dynamically to make better group recommendations. However, it does not consider group member interactions with different recommendation items compared with AGREE, where AGREE can learn group embeddings by using one neural attention mechanism for the group member-item interactions. Besides, it has been observed that when user and item embeddings are trained with the deep reinforcement learning framework directly together, the experiment result can be unstable and lead to a bad performance.

In order to improve the DRGR performances, an updated attention network can be used in the current state embedding layer. After embedding members in one group, we can use one neural attention mechanism to combine the group member embeddings and history item embeddings then generate one new state embedding. This neural attention can allow different group members have different influences when the group has various item histories. So the DRGR model would have more flexibility when items are diverse. In addition, we can incorporate the Deep Sets \cite{zaheer2018deep} ideas to aggregate group member state embeddings. Each group member embedding will consist its user preference embedding and its browsing history item embeddings. Moreover, we can compare our model with baselines on more datasets, such as the tourist attractions recommendation so that we can utilize their group information.

\begin{acks}

We would like to extend our sincere thanks to Dr. Srijan Kumar for his valuable suggestions.

\end{acks}

\bibliographystyle{ACM-Reference-Format}
\bibliography{reference}


\begin{thebibliography}{18}


\ifx \showCODEN    \undefined \def \showCODEN     #1{\unskip}     \fi
\ifx \showDOI      \undefined \def \showDOI       #1{#1}\fi
\ifx \showISBNx    \undefined \def \showISBNx     #1{\unskip}     \fi
\ifx \showISBNxiii \undefined \def \showISBNxiii  #1{\unskip}     \fi
\ifx \showISSN     \undefined \def \showISSN      #1{\unskip}     \fi
\ifx \showLCCN     \undefined \def \showLCCN      #1{\unskip}     \fi
\ifx \shownote     \undefined \def \shownote      #1{#1}          \fi
\ifx \showarticletitle \undefined \def \showarticletitle #1{#1}   \fi
\ifx \showURL      \undefined \def \showURL       {\relax}        \fi
\providecommand\bibfield[2]{#2}
\providecommand\bibinfo[2]{#2}
\providecommand\natexlab[1]{#1}
\providecommand\showeprint[2][]{arXiv:#2}

\bibitem[\protect\citeauthoryear{Cao, He, Miao, An, Yang, and Hong}{Cao
  et~al\mbox{.}}{2018}]%
        {cao2018attentive}
\bibfield{author}{\bibinfo{person}{Da Cao}, \bibinfo{person}{Xiangnan He},
  \bibinfo{person}{Lianhai Miao}, \bibinfo{person}{Yahui An},
  \bibinfo{person}{Chao Yang}, {and} \bibinfo{person}{Richang Hong}.}
  \bibinfo{year}{2018}\natexlab{}.
\newblock \showarticletitle{Attentive Group Recommendation}. In
  \bibinfo{booktitle}{\emph{The 41st International ACM SIGIR Conference on
  Research \& Development in Information Retrieval}} (Ann Arbor, MI, USA)
  \emph{(\bibinfo{series}{SIGIR '18})}. \bibinfo{publisher}{Association for
  Computing Machinery}, \bibinfo{address}{New York, NY, USA},
  \bibinfo{pages}{645–654}.
\newblock
\showISBNx{9781450356572}
\urldef\tempurl%
\url{https://doi.org/10.1145/3209978.3209998}
\showDOI{\tempurl}


\bibitem[\protect\citeauthoryear{Chen, Yu, Da, Tan, Huang, and Tang}{Chen
  et~al\mbox{.}}{2018}]%
        {chen2018stabilizing}
\bibfield{author}{\bibinfo{person}{Shi-Yong Chen}, \bibinfo{person}{Yang Yu},
  \bibinfo{person}{Qing Da}, \bibinfo{person}{Jun Tan},
  \bibinfo{person}{Hai-Kuan Huang}, {and} \bibinfo{person}{Hai-Hong Tang}.}
  \bibinfo{year}{2018}\natexlab{}.
\newblock \showarticletitle{Stabilizing Reinforcement Learning in Dynamic
  Environment with Application to Online Recommendation}. In
  \bibinfo{booktitle}{\emph{Proceedings of the 24th ACM SIGKDD International
  Conference on Knowledge Discovery \& Data Mining}} (London, United Kingdom)
  \emph{(\bibinfo{series}{KDD '18})}. \bibinfo{publisher}{Association for
  Computing Machinery}, \bibinfo{address}{New York, NY, USA},
  \bibinfo{pages}{1187–1196}.
\newblock
\showISBNx{9781450355520}
\urldef\tempurl%
\url{https://doi.org/10.1145/3219819.3220122}
\showDOI{\tempurl}


\bibitem[\protect\citeauthoryear{Cheng, Koc, Harmsen, Shaked, Chandra, Aradhye,
  Anderson, Corrado, Chai, Ispir, Anil, Haque, Hong, Jain, Liu, and Shah}{Cheng
  et~al\mbox{.}}{2016}]%
        {cheng2016wide}
\bibfield{author}{\bibinfo{person}{Heng-Tze Cheng}, \bibinfo{person}{Levent
  Koc}, \bibinfo{person}{Jeremiah Harmsen}, \bibinfo{person}{Tal Shaked},
  \bibinfo{person}{Tushar Chandra}, \bibinfo{person}{Hrishi Aradhye},
  \bibinfo{person}{Glen Anderson}, \bibinfo{person}{Greg Corrado},
  \bibinfo{person}{Wei Chai}, \bibinfo{person}{Mustafa Ispir},
  \bibinfo{person}{Rohan Anil}, \bibinfo{person}{Zakaria Haque},
  \bibinfo{person}{Lichan Hong}, \bibinfo{person}{Vihan Jain},
  \bibinfo{person}{Xiaobing Liu}, {and} \bibinfo{person}{Hemal Shah}.}
  \bibinfo{year}{2016}\natexlab{}.
\newblock \bibinfo{title}{Wide \& Deep Learning for Recommender Systems}.
\newblock
\newblock
\showeprint[arxiv]{1606.07792}~[cs.LG]


\bibitem[\protect\citeauthoryear{Choi, Ha, Hwang, Kim, Ha, and Yoon}{Choi
  et~al\mbox{.}}{2018}]%
        {choi2018reinforcement}
\bibfield{author}{\bibinfo{person}{Sungwoon Choi}, \bibinfo{person}{Heonseok
  Ha}, \bibinfo{person}{Uiwon Hwang}, \bibinfo{person}{Chanju Kim},
  \bibinfo{person}{Jung-Woo Ha}, {and} \bibinfo{person}{Sungroh Yoon}.}
  \bibinfo{year}{2018}\natexlab{}.
\newblock \bibinfo{title}{Reinforcement Learning based Recommender System using
  Biclustering Technique}.
\newblock
\newblock
\showeprint[arxiv]{1801.05532}~[cs.IR]


\bibitem[\protect\citeauthoryear{Harper and Konstan}{Harper and
  Konstan}{2015}]%
        {harper2015movielens}
\bibfield{author}{\bibinfo{person}{F.~Maxwell Harper} {and}
  \bibinfo{person}{Joseph~A. Konstan}.} \bibinfo{year}{2015}\natexlab{}.
\newblock \showarticletitle{The MovieLens Datasets: History and Context}.
\newblock \bibinfo{journal}{\emph{ACM Trans. Interact. Intell. Syst.}}
  \bibinfo{volume}{5}, \bibinfo{number}{4}, Article \bibinfo{articleno}{19}
  (\bibinfo{date}{Dec.} \bibinfo{year}{2015}), \bibinfo{numpages}{19}~pages.
\newblock
\showISSN{2160-6455}
\urldef\tempurl%
\url{https://doi.org/10.1145/2827872}
\showDOI{\tempurl}


\bibitem[\protect\citeauthoryear{Liu, Tang, Li, Zhang, Ye, Chen, Guo, and
  Zhang}{Liu et~al\mbox{.}}{2019}]%
        {liu2019deep}
\bibfield{author}{\bibinfo{person}{Feng Liu}, \bibinfo{person}{Ruiming Tang},
  \bibinfo{person}{Xutao Li}, \bibinfo{person}{Weinan Zhang},
  \bibinfo{person}{Yunming Ye}, \bibinfo{person}{Haokun Chen},
  \bibinfo{person}{Huifeng Guo}, {and} \bibinfo{person}{Yuzhou Zhang}.}
  \bibinfo{year}{2019}\natexlab{}.
\newblock \bibinfo{title}{Deep Reinforcement Learning based Recommendation with
  Explicit User-Item Interactions Modeling}.
\newblock
\newblock
\showeprint[arxiv]{1810.12027}~[cs.IR]


\bibitem[\protect\citeauthoryear{{Munemasa}, {Tomomatsu}, {Hayashi}, and
  {Takagi}}{{Munemasa} et~al\mbox{.}}{2018}]%
        {munemasa2018deep}
\bibfield{author}{\bibinfo{person}{I. {Munemasa}}, \bibinfo{person}{Y.
  {Tomomatsu}}, \bibinfo{person}{K. {Hayashi}}, {and} \bibinfo{person}{T.
  {Takagi}}.} \bibinfo{year}{2018}\natexlab{}.
\newblock \showarticletitle{Deep reinforcement learning for recommender
  systems}. In \bibinfo{booktitle}{\emph{2018 International Conference on
  Information and Communications Technology (ICOIACT)}}.
  \bibinfo{pages}{226--233}.
\newblock
\urldef\tempurl%
\url{https://doi.org/10.1109/ICOIACT.2018.8350761}
\showDOI{\tempurl}


\bibitem[\protect\citeauthoryear{Paszke, Gross, Massa, Lerer, Bradbury, Chanan,
  Killeen, Lin, Gimelshein, Antiga, Desmaison, Kopf, Yang, DeVito, Raison,
  Tejani, Chilamkurthy, Steiner, Fang, Bai, and Chintala}{Paszke
  et~al\mbox{.}}{2019}]%
        {adam2019pytorch}
\bibfield{author}{\bibinfo{person}{Adam Paszke}, \bibinfo{person}{Sam Gross},
  \bibinfo{person}{Francisco Massa}, \bibinfo{person}{Adam Lerer},
  \bibinfo{person}{James Bradbury}, \bibinfo{person}{Gregory Chanan},
  \bibinfo{person}{Trevor Killeen}, \bibinfo{person}{Zeming Lin},
  \bibinfo{person}{Natalia Gimelshein}, \bibinfo{person}{Luca Antiga},
  \bibinfo{person}{Alban Desmaison}, \bibinfo{person}{Andreas Kopf},
  \bibinfo{person}{Edward Yang}, \bibinfo{person}{Zachary DeVito},
  \bibinfo{person}{Martin Raison}, \bibinfo{person}{Alykhan Tejani},
  \bibinfo{person}{Sasank Chilamkurthy}, \bibinfo{person}{Benoit Steiner},
  \bibinfo{person}{Lu Fang}, \bibinfo{person}{Junjie Bai}, {and}
  \bibinfo{person}{Soumith Chintala}.} \bibinfo{year}{2019}\natexlab{}.
\newblock \showarticletitle{PyTorch: An Imperative Style, High-Performance Deep
  Learning Library}.
\newblock In \bibinfo{booktitle}{\emph{Advances in Neural Information
  Processing Systems 32}}, \bibfield{editor}{\bibinfo{person}{H.~Wallach},
  \bibinfo{person}{H.~Larochelle}, \bibinfo{person}{A.~Beygelzimer},
  \bibinfo{person}{F.~d\textquotesingle Alch\'{e}-Buc},
  \bibinfo{person}{E.~Fox}, {and} \bibinfo{person}{R.~Garnett}} (Eds.).
  \bibinfo{publisher}{Curran Associates, Inc.}, \bibinfo{pages}{8024--8035}.
\newblock
\urldef\tempurl%
\url{http://papers.neurips.cc/paper/9015-pytorch-an-imperative-style-high-performance-deep-learning-library.pdf}
\showURL{%
\tempurl}


\bibitem[\protect\citeauthoryear{Sankar, Wu, Wu, Zhang, Yang, and
  Sundaram}{Sankar et~al\mbox{.}}{2020}]%
        {sankar2020groupim}
\bibfield{author}{\bibinfo{person}{Aravind Sankar}, \bibinfo{person}{Yanhong
  Wu}, \bibinfo{person}{Yuhang Wu}, \bibinfo{person}{Wei Zhang},
  \bibinfo{person}{Hao Yang}, {and} \bibinfo{person}{Hari Sundaram}.}
  \bibinfo{year}{2020}\natexlab{}.
\newblock \showarticletitle{GroupIM: A Mutual Information Maximization
  Framework for Neural Group Recommendation}. In
  \bibinfo{booktitle}{\emph{Proceedings of the 43rd International ACM SIGIR
  Conference on Research and Development in Information Retrieval}}.
  \bibinfo{pages}{1279--1288}.
\newblock
\urldef\tempurl%
\url{https://doi.org/10.1145/3397271.3401116}
\showDOI{\tempurl}


\bibitem[\protect\citeauthoryear{Shan, Hoens, Jiao, Wang, Yu, and Mao}{Shan
  et~al\mbox{.}}{2016}]%
        {ying2016deep}
\bibfield{author}{\bibinfo{person}{Ying Shan}, \bibinfo{person}{T.~Ryan Hoens},
  \bibinfo{person}{Jian Jiao}, \bibinfo{person}{Haijing Wang},
  \bibinfo{person}{Dong Yu}, {and} \bibinfo{person}{JC Mao}.}
  \bibinfo{year}{2016}\natexlab{}.
\newblock \showarticletitle{Deep Crossing: Web-Scale Modeling without Manually
  Crafted Combinatorial Features}. In \bibinfo{booktitle}{\emph{Proceedings of
  the 22nd ACM SIGKDD International Conference on Knowledge Discovery and Data
  Mining}} (San Francisco, California, USA) \emph{(\bibinfo{series}{KDD '16})}.
  \bibinfo{publisher}{Association for Computing Machinery},
  \bibinfo{address}{New York, NY, USA}, \bibinfo{pages}{255–262}.
\newblock
\showISBNx{9781450342322}
\urldef\tempurl%
\url{https://doi.org/10.1145/2939672.2939704}
\showDOI{\tempurl}


\bibitem[\protect\citeauthoryear{Vinh~Tran, Nguyen~Pham, Tay, Liu, Cong, and
  Li}{Vinh~Tran et~al\mbox{.}}{2019}]%
        {tran2019interact}
\bibfield{author}{\bibinfo{person}{Lucas Vinh~Tran}, \bibinfo{person}{Tuan-Anh
  Nguyen~Pham}, \bibinfo{person}{Yi Tay}, \bibinfo{person}{Yiding Liu},
  \bibinfo{person}{Gao Cong}, {and} \bibinfo{person}{Xiaoli Li}.}
  \bibinfo{year}{2019}\natexlab{}.
\newblock \showarticletitle{Interact and Decide: Medley of Sub-Attention
  Networks for Effective Group Recommendation}. In
  \bibinfo{booktitle}{\emph{Proceedings of the 42nd International ACM SIGIR
  Conference on Research and Development in Information Retrieval}} (Paris,
  France) \emph{(\bibinfo{series}{SIGIR'19})}. \bibinfo{publisher}{Association
  for Computing Machinery}, \bibinfo{address}{New York, NY, USA},
  \bibinfo{pages}{255–264}.
\newblock
\showISBNx{9781450361729}
\urldef\tempurl%
\url{https://doi.org/10.1145/3331184.3331251}
\showDOI{\tempurl}


\bibitem[\protect\citeauthoryear{Wang, Wang, Hsu, and Wang}{Wang
  et~al\mbox{.}}{2014}]%
        {wang2014exploration}
\bibfield{author}{\bibinfo{person}{Xinxi Wang}, \bibinfo{person}{Yi Wang},
  \bibinfo{person}{David Hsu}, {and} \bibinfo{person}{Ye Wang}.}
  \bibinfo{year}{2014}\natexlab{}.
\newblock \showarticletitle{Exploration in Interactive Personalized Music
  Recommendation: A Reinforcement Learning Approach}.
\newblock \bibinfo{journal}{\emph{ACM Trans. Multimedia Comput. Commun. Appl.}}
  \bibinfo{volume}{11}, \bibinfo{number}{1}, Article \bibinfo{articleno}{7}
  (\bibinfo{date}{Sept.} \bibinfo{year}{2014}), \bibinfo{numpages}{22}~pages.
\newblock
\showISSN{1551-6857}
\urldef\tempurl%
\url{https://doi.org/10.1145/2623372}
\showDOI{\tempurl}


\bibitem[\protect\citeauthoryear{Zaheer, Kottur, Ravanbakhsh, Poczos,
  Salakhutdinov, and Smola}{Zaheer et~al\mbox{.}}{2017}]%
        {zaheer2018deep}
\bibfield{author}{\bibinfo{person}{Manzil Zaheer}, \bibinfo{person}{Satwik
  Kottur}, \bibinfo{person}{Siamak Ravanbakhsh}, \bibinfo{person}{Barnabas
  Poczos}, \bibinfo{person}{Russ~R Salakhutdinov}, {and}
  \bibinfo{person}{Alexander~J Smola}.} \bibinfo{year}{2017}\natexlab{}.
\newblock \showarticletitle{Deep Sets}. In \bibinfo{booktitle}{\emph{Advances
  in Neural Information Processing Systems}},
  \bibfield{editor}{\bibinfo{person}{I.~Guyon}, \bibinfo{person}{U.~V.
  Luxburg}, \bibinfo{person}{S.~Bengio}, \bibinfo{person}{H.~Wallach},
  \bibinfo{person}{R.~Fergus}, \bibinfo{person}{S.~Vishwanathan}, {and}
  \bibinfo{person}{R.~Garnett}} (Eds.), Vol.~\bibinfo{volume}{30}.
  \bibinfo{publisher}{Curran Associates, Inc.}
\newblock
\urldef\tempurl%
\url{https://proceedings.neurips.cc/paper/2017/file/f22e4747da1aa27e363d86d40ff442fe-Paper.pdf}
\showURL{%
\tempurl}


\bibitem[\protect\citeauthoryear{Zhang, Yao, Sun, and Tay}{Zhang
  et~al\mbox{.}}{2019}]%
        {zhang2019deep}
\bibfield{author}{\bibinfo{person}{Shuai Zhang}, \bibinfo{person}{Lina Yao},
  \bibinfo{person}{Aixin Sun}, {and} \bibinfo{person}{Yi Tay}.}
  \bibinfo{year}{2019}\natexlab{}.
\newblock \showarticletitle{Deep Learning Based Recommender System}.
\newblock \bibinfo{journal}{\emph{Comput. Surveys}} \bibinfo{volume}{52},
  \bibinfo{number}{1} (\bibinfo{date}{Feb} \bibinfo{year}{2019}),
  \bibinfo{pages}{1–38}.
\newblock
\showISSN{1557-7341}
\urldef\tempurl%
\url{https://doi.org/10.1145/3285029}
\showDOI{\tempurl}


\bibitem[\protect\citeauthoryear{Zhao, Xia, Zhang, Ding, Yin, and Tang}{Zhao
  et~al\mbox{.}}{2018a}]%
        {zhao2018deep}
\bibfield{author}{\bibinfo{person}{Xiangyu Zhao}, \bibinfo{person}{Long Xia},
  \bibinfo{person}{Liang Zhang}, \bibinfo{person}{Zhuoye Ding},
  \bibinfo{person}{Dawei Yin}, {and} \bibinfo{person}{Jiliang Tang}.}
  \bibinfo{year}{2018}\natexlab{a}.
\newblock \showarticletitle{Deep reinforcement learning for page-wise
  recommendations}.
\newblock \bibinfo{journal}{\emph{Proceedings of the 12th ACM Conference on
  Recommender Systems}} (\bibinfo{date}{Sep} \bibinfo{year}{2018}).
\newblock
\showISBNx{9781450359016}
\urldef\tempurl%
\url{https://doi.org/10.1145/3240323.3240374}
\showDOI{\tempurl}


\bibitem[\protect\citeauthoryear{Zhao, Zhang, Ding, Xia, Tang, and Yin}{Zhao
  et~al\mbox{.}}{2018b}]%
        {zhao2018recommendations}
\bibfield{author}{\bibinfo{person}{Xiangyu Zhao}, \bibinfo{person}{Liang
  Zhang}, \bibinfo{person}{Zhuoye Ding}, \bibinfo{person}{Long Xia},
  \bibinfo{person}{Jiliang Tang}, {and} \bibinfo{person}{Dawei Yin}.}
  \bibinfo{year}{2018}\natexlab{b}.
\newblock \showarticletitle{Recommendations with Negative Feedback via Pairwise
  Deep Reinforcement Learning}.
\newblock \bibinfo{journal}{\emph{Proceedings of the 24th ACM SIGKDD
  International Conference on Knowledge Discovery \& Data Mining}}
  (\bibinfo{date}{Jul} \bibinfo{year}{2018}).
\newblock
\showISBNx{9781450355520}
\urldef\tempurl%
\url{https://doi.org/10.1145/3219819.3219886}
\showDOI{\tempurl}


\bibitem[\protect\citeauthoryear{Zhao, Zhang, Xia, Ding, Yin, and Tang}{Zhao
  et~al\mbox{.}}{2019}]%
        {zhao2019deep}
\bibfield{author}{\bibinfo{person}{Xiangyu Zhao}, \bibinfo{person}{Liang
  Zhang}, \bibinfo{person}{Long Xia}, \bibinfo{person}{Zhuoye Ding},
  \bibinfo{person}{Dawei Yin}, {and} \bibinfo{person}{Jiliang Tang}.}
  \bibinfo{year}{2019}\natexlab{}.
\newblock \bibinfo{title}{Deep Reinforcement Learning for List-wise
  Recommendations}.
\newblock
\newblock
\showeprint[arxiv]{1801.00209}~[cs.LG]


\bibitem[\protect\citeauthoryear{Zheng, Zhang, Zheng, Xiang, Yuan, Xie, and
  Li}{Zheng et~al\mbox{.}}{2018}]%
        {zheng2018drn}
\bibfield{author}{\bibinfo{person}{Guanjie Zheng}, \bibinfo{person}{Fuzheng
  Zhang}, \bibinfo{person}{Zihan Zheng}, \bibinfo{person}{Yang Xiang},
  \bibinfo{person}{Nicholas~Jing Yuan}, \bibinfo{person}{Xing Xie}, {and}
  \bibinfo{person}{Zhenhui Li}.} \bibinfo{year}{2018}\natexlab{}.
\newblock \showarticletitle{DRN: A Deep Reinforcement Learning Framework for
  News Recommendation}. In \bibinfo{booktitle}{\emph{Proceedings of the 2018
  World Wide Web Conference}} (Lyon, France) \emph{(\bibinfo{series}{WWW
  '18})}. \bibinfo{publisher}{International World Wide Web Conferences Steering
  Committee}, \bibinfo{address}{Republic and Canton of Geneva, CHE},
  \bibinfo{pages}{167–176}.
\newblock
\showISBNx{9781450356398}
\urldef\tempurl%
\url{https://doi.org/10.1145/3178876.3185994}
\showDOI{\tempurl}


\end{thebibliography}




\end{document}